\documentclass[9pt]{article}

\usepackage{amsthm}
\usepackage{dcolumn}
\usepackage{amsmath}
\usepackage{bm}
\usepackage{graphicx}
\usepackage{xcolor}
\usepackage{ulem}

\newtheorem{thm}{Theorem}[section]
\newtheorem{cor}{Corollary}[section]
\newtheorem{lem}{Lemma}[section]

\newtheorem{prop}{Proposition}

\input{Qcircuit}

\newcommand{\op}[2]{|#1\rangle \langle #2|}

\newcommand {\diag } {{\rm diag}}

\title{Optimal simulation of three-qubit gates}

\author{Nengkun Yu and Mingsheng Ying\\
\textit{State Key Laboratory of Intelligent Technology and Systems,}\\
\textit{Tsinghua National Laboratory for Information Science and Technology,}\\
\textit{Department of Computer Science and Technology,}\\
\textit{Tsinghua University, Beijing 100084, China}\\
\textit{Center for Quantum Computation and Intelligent Systems (QCIS),}\\
\textit{Faculty of Engineering and Information Technology, }\\
\textit{University of Technology, Sydney, NSW 2007, Australia}
}
\begin{document}
\maketitle
\begin{abstract}
In this paper, we study the optimal simulation of three-qubit unitary by using two-qubit gates. First, we give a lower bound on the two-qubit gates cost of simulating a multi-qubit gate. Secondly, we completely characterize the two-qubit gate cost of simulating a three-qubit controlled controlled gate by generalizing our result on the cost of Toffoli gate. The function of controlled controlled gate is simply a three-qubit controlled unitary gate and can be intuitively explained as follows: the gate will output the states of the two control qubit directly, and apply the given one-qubit unitary $u$ on the target qubit only if both the states of the control are $\ket{1}$. Previously, it is only known that five two-qubit gates is sufficient for implementing such a gate [Sleator and Weinfurter, Phys. Rev. Lett. 74, 4087 (1995)]. Our result shows that if the determinant of $u$ is 1, four two-qubit gates is achievable optimal. Otherwise, five is optimal. Thirdly, we show that five two-qubit gates are necessary and sufficient for implementing the Fredkin gate(the controlled swap gate), which settles the open problem introduced in [Smolin and DiVincenzo, Phys. Rev. A, 53, 2855 (1996)].
The Fredkin gate is one of the most important quantum logic gates because it is universal alone for classical reversible computation, and thus with little help, universal for quantum computation. Before our work, a five two-qubit gates decomposition of the Fredkin gate was already known, and numerical evidence of showing five is optimal is found.
\end{abstract}

\section{Introduction}A fundamental
issue of several interacting systems is to quantify the strength of this interaction. Particularly valuable are techniques that can
compare interactions of quite different types of system or
particles and different physical manifestations of the interaction.
 In order to provide an intellectual
background of comparing interaction strength of different physical systems, robust characterizations
is desirable.
Quantum information theory has provided quite new insights
into this question. In particular, there has been considerable progress in
quantifying the strength of Hamiltonian and unitary interactions \cite{DVC+01,CLS01,EJPP00,CDKL01,BS03,BHLS03,LSW09,KC01,YDY10} for bipartite system. The starting point was the theory of entanglement
of quantum states which quantifies how
much nonclassical correlation the state embodies. It would be quite interesting to extend these results to multipartite system. The most intriguing approach might be to study how much bipartite correlation is needed to implement a multipartite correlation. In other words, how many two-qubit unitary is needed to simulate a given multi-qubit gate? One obstacle standing in front of such desire is the fact that multipartite entanglement is difficult to characterize. A possible direction is to consider those symmetric gates.

This fundamental topic is clearly of interest to experimentalists who try to
create systems in interaction. A great challenge in the contemporary science and engineering is building a full-fledged quantum computer, which is essentially a large quantum circuit consisting of basic quantum logical gates. In order to accomplish a quantum algorithm, even in a small size, one has to implement a relatively high level of control over the multi-qubit quantum system. It has also been experimentally demonstrated that two-qubit gates can be realized with high fidelity using the current technology, for example, two-qubit gate with superconducting quibts have been presented with fidelities higher than $90\%$ \cite{DCG+09}. Finding more efficient ways to implement quantum gates may allow small-scale quantum computing
tasks to be demonstrated on a shorter time scale.

Due to its significance in quantum computing, lots of efforts have been devoted to study correlation of controlled unitary, see \cite{NC00,DIV98,DS94,SW95,BBCD+95} as a quite incomplete list. But no affirm result is known, even for some highly symmetric three-qubit gate.
Very recently, we have showed that five two-qubit gates are optimal for implementing a Toffoli gate \cite{YDY13} by employing basic techniques from
quantum information.

Another gate that has received particular attention is the three qubit
conditional swap gate, or Fredkin gate. The Fredkin gate
is of interest because it is a universal gate for classical reversible
computation\cite{FT82}, which means that any logical or arithmetic operation can be constructed entirely of Fredkin gates. The quantum version has been
used by Ekert and Macchiavello to design a circuit for
error correcting quantum computations with the symmetric subspace
method of \cite{BDJ94}. The experimental and theoretical pursuit of efficient implementation of the Fredkin gate using a sequence of single- and two-qubit gates has quite a long history. A simple optical model to realize a reversible, potentially error-free logic gate¡ªa Fredkin gate is proposed in \cite{MIL89}.  Chau and Wilczek give a specific six-gate construction
of the Fredkin
gate in \cite{CW95}. An analytic five-gate construction is presented
and numerical tests suggest is minimal in \cite{SD96}.

In this paper, the technique introduced in \cite{YDY13} is used to deal with generalized three-qubit controlled controlled gate and Fredkin gate. The two-qubit gates cost on simulating three-qubit controlled controlled gate is completely characterized. More precisely, it is showed that any controlled controlled gate requires at least four two-qubit gates to simulate. If the determinant of the controlled unitary is one, then four gates simulation is achievable. Otherwise, five gates implementation is optimal. Later, we present a theoretical proof that a five two-bit gate is indeed the optimal implementation of the Fredkin gate.
\section{Preliminaries and Notations}

Note that any bipartite unitary $U_{AB}$ acting on a qubit system $A$ and a general system $B$ is said to be a controlled-gate with control on $A$ if it can be decomposed into the form of $$U_{AB}=\op{0_A}{0_A}\otimes U_0+\op{1_A}{1_A}\otimes U_1.$$

A controlled-controlled gate, acting on three quantum bits, namely  $A$,$B$, and $C$. Here $A$ and $B$ are control qubits, and $C$ is the target qubit with computational basis $\{\ket{0},\ket{1}\}$ for each qubit. Upon input $\ket{abc}$, the gate will output the states of $A$ and $B$ directly, and apply $U$ on the system $C$ only if both the states of $A$ and $B$ are $\ket{1}$.

In this letter, each three-qubit gate is regarded as a unitary transformation performed on a tripartite system $ABC$, all the two-qubit gates employed to implement three-qubit gate can be simply classified into three classes: class $\mathcal{K}_{AB}$ - the gates acting on the subsystem $AB$,  class $\mathcal{K}_{BC}$ - the gates on $BC$, and class $\mathcal{K}_{AC}$ - the gates on $AC$. Obviously, it is impossible that all the two-qubit gates used to simulate $F_{ABC}$ belong to a single one of the three classes $\mathcal{K}_{AB}$, $\mathcal{K}_{BC}$, $\mathcal{K}_{AC}$.

The validity of following propositions is showed in \cite{YDY13}.
\begin{prop}
Any two-dimensional $2\otimes 2$ state subspace contains some product state.
\end{prop}
\begin{prop}
If $U_{AB}U_{AC}$ is a three-qubit controlled unitary with control on $A$, where $U_{AB}\in \mathcal{K}_{AB}$ and $U_{AC}\in\mathcal{K}_{AC}$,
then there exist $v_{B1},v_{B2}$ and $w_{C1},w_{C2}$ being one-qubit unitaries on $\mathcal{H}_B$ and $\mathcal{H}_C$ such that \[U_{AB}U_{AC}=\op{0}{0}\otimes v_{B1}\otimes w_{C1}+\op{1}{1}\otimes v_{B2}\otimes w_{C2}.\]
\end{prop}

\section{Lower Bound for simulating general multi-qubit gate}

First, we give a general bound on the cost on the two-qubit gates for implementing milti-qubit gates.
\begin{thm}
Almost any $n$-qubit gate requires at least $\lceil\frac{4^n-3n-1}{9}\rceil$ two-qubit unitaries to implement without ancilla.
\end{thm}
\textit{Proof---}: This theorem is proved by simply counting the degree of freedom(DOF).

Without loss of generality(wlog), we only consider quantum gates with unit determinant, $i.e,$ $\det U=1$. It is well known that the DOF of $n-$qubit unitary with unit determinant is $4^n-1$. As a special case, the DOF of such two-qubit unitary becomes 15. More precisely, it is proved in \cite{KC01} that each two-qubit unitary operation $U$ can be expressed
into form:
$U=(u_A\otimes u_B)U_d(v_A\otimes v_B),$
where $u_A$, $u_B$, $v_A$, $v_B$ are one-qubit unitary gates with unit determinant,
and $U_d=\exp[i(\alpha_x X\otimes X+\alpha_y
Y\otimes Y+\alpha_z Z\otimes Z)]$, and
$X$, $Y$, $Z$ are Pauli matrices.

Now we assume that arbitrary $n$-qubit gates of system $\mathcal{H}_{A1}\otimes \mathcal{H}_{A2}\otimes \cdots\mathcal{H}_{A_n}$ could be implemented by some circuit consisting of $k$ two-qubit gates. Notice that the structure of the circuit with five two-qubit gates is finite. For any fixed structure, there are $n_i$ gates on $\mathcal{H}_{Ai}$, respectively. It is easy to see that $\sum_{i=1}^n n_i=2k$. There are $n_i+1$ local unitaries on $\mathcal{H}_{Ai}$, the DOF for this part is $3(n_i+1)$.  Noticing the DOF of each two-qubit gate is $3$, then the DOF of the whole circuit is less or equal to
\[
3\times k+3\times(\sum_{i=1}^n (n_i+1))=9k+3n.
\]
It is easy to see that
\[
9k+3n\geq 4^n-1\Rightarrow k\geq\lceil\frac{4^n-3n-1}{9}\rceil
\]
The proof of this theorem is complete. 

Let $n=3$, we know that the simulation of a general three-qubit gate would require at least $\lceil\frac{4^3-3\times 3-1}{9}\rceil=6$ two-qubit gates. The following nature question arises,

\textit{Whether six two-qubit quantum gates are sufficient to generate any three-bit quantum gate?}

Unfortunately, we are not able to solve this problem. Instead, we can show the optimal implementation of some symmetric three-qubit gates in the following sections.

\section{Optimal simulation of controlled-controlled gate }

The particular ``controlled-controlled" gate is introduced by Deutsch in \cite{DEU89} and it is proved that some of such gate is universal for the first time. That is they are adequate for constructing networks with any possible quantum computational property.

It is proved in \cite{SW95} that any controlled-controlled gate can be implemented by using only five two-qubit gates circuit,
 
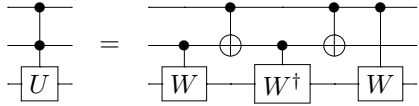
\[
\Qcircuit @C=.5em @R=0em @!R {
& \ctrl{1} & \qw & & & \qw & \ctrl{1} & \qw & \ctrl{1} & \ctrl{2} & \qw\\
& \ctrl{1} & \qw & \push{\rule{.3em}{0em}=\rule{.3em}{0em}} & & \ctrl{1} & \targ & \ctrl{1} & \targ & \qw & \qw\\
& \gate{U} & \qw & & & \gate{W} & \qw & \gate{W^\dag} & \qw & \gate{W} & \qw
}
\]
The left circuit denotes ``controlled-controlled-$U$" gate, $W$ on the right hand side is a unitary satisfying $W^2=U$.

To study the two-qubit gate cost for implementing the general ``controlled-controlled-$U$" gate, we only need to deal with diagonal $U$, $i.e.$,
\begin{eqnarray*}
V(\theta_1,\theta_2)=\left(
\begin{array}{cccccccc}
 1 & 0 & 0 & 0 & 0 & 0 & 0& 0\\
 0 & 1 & 0 & 0 & 0 & 0 & 0& 0\\
 0 & 0 & 1 & 0 & 0 & 0 & 0& 0\\
 0 & 0 & 0 & 1 & 0 & 0 & 0& 0\\
 0 & 0 & 0 & 0 & 1 & 0 & 0& 0\\
 0 & 0 & 0 & 0 & 0 & 1 & 0& 0\\
 0 & 0 & 0 & 0 & 0 & 0 & e^{i\theta_1}& 0\\
 0 & 0 & 0 & 0 & 0 & 0 & 0& e^{i\theta_2}\\
\end{array}
\right).
\end{eqnarray*}

$V(\theta_1,\theta_2)_{ABC}$ is regarded as a tripartite unitary with control on $A$ and $B$. It can be considered as a control unitary with control on $A$, $B$ or $C$,
$$V(\theta_1,\theta_2)_{ABC}=\op{0}{0}\otimes I_{BC}+\op{1}{1}\otimes R_{BC},$$
where $R=\op{00}{00}+\op{01}{01}+e^{i\theta_1}\op{10}{10}+e^{i\theta_2}\op{11}{11}$.

Also, we can verify that $V(\theta_1,\theta_2)_{ABC}$ is invariant under the permutation of $A$ and $B$, that is $V(\theta_1,\theta_2)_{ABC}=S_{AB}V(\theta_1,\theta_2)_{ABC}S_{AB}$ with $S_{BC}$ denoting the swap gate on $A$ and $B$. These observations are used during the following argument.

The following result from \cite{YDY13} is useful for our discussion.
\begin{thm}
$V(0,\theta)=I-(1-e^{i\theta})\op{111}{111}$ requires five two-qubit gates to simulate, provide that $e^{i\theta}\neq 1$.
\end{thm}
One can easily verify the following equation
\begin{eqnarray*}
V(0,\theta_2-\theta_1)_{ABC}=V(\theta_1,\theta_2)_{ABC}W(-\theta_1)_{AB}=W(-\theta_1)_{AB}V(\theta_1,\theta_2)_{ABC},
\end{eqnarray*}
where $W(\theta)=\op{00}{00}+\op{01}{01}+\op{10}{10}+e^{i\theta}\op{11}{11}$.
Therefore, we can conclude that
\begin{lem}
Any three qubit controlled controlled gate would require at least four two-qubit gates to simulate.
\end{lem}
Suppose $e^{i\theta_1}\neq e^{i\theta_2}$. If such $V(\theta_1,\theta_2)$ can be implemented by three or less two-qubit gates, then four two-qubit gates or less can simulate some $V(0,\theta)=I-(1-e^{i\theta})\op{111}{111}$, conflict from the above theorem.

For $V(-\theta,\theta)$, we can find the following simulation circuit consisting of fout two-qubit gates, therefore, it is optimal.
 
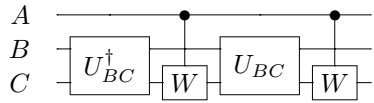
\[
\Qcircuit @C=.5em @R=0em @!R {
\lstick{A} &&  \qw                          & \ctrl{2}     & \qw                   & \ctrl{2}&\qw\\
\lstick{B} &&  \multigate{1}{U_{BC}^{\dag}} &  \qw         & \multigate{1}{U_{BC}} & \qw &\qw\\
\lstick{C} &&  \ghost{U_{BC}^{\dag}}        & \gate{W}     & \ghost{U_{BC}}        & \gate{W}  &\qw
}
\]
where
\begin{eqnarray*}
W=\left(
\begin{array}{cc}
e^{-i\theta/2} & 0 \\
0 & e^{i\theta/2}
\end{array}
\right).
\end{eqnarray*}
and $U_{BC}(W_C\otimes I_B)U_{BC}^{\dag}=\diag\{e^{i\theta/2},e^{-i\theta/2},e^{-i\theta/2},e^{i\theta/2}\}$. Such $U_{BC}$ does exist since the eigenvalues of $W_C\otimes I_B$ are $\{e^{i\theta/2},e^{-i\theta/2},e^{-i\theta/2},e^{i\theta/2}\}$.

In order to study the rest case, we begin from the following special circuit.
\begin{lem}
There is no $U_{AC},V_{AC}\in \mathcal{K}_{AC}$ and $U_{BC},V_{BC}\in \mathcal{K}_{BC}$ such that $V(\theta_1,\theta_2)$ can be implemented in the following circuit with $e^{i(\theta_1+\theta_2)}\neq 1$ and $e^{i\theta_1}\neq e^{i\theta_2}$,
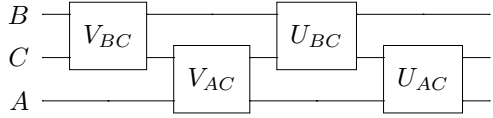
\[
\Qcircuit @C=1em @R=.7em {
\lstick{B} & \multigate{1}{V_{BC}} & \qw & \multigate{1}{U_{BC}} & \qw &\qw\\
\lstick{C} & \ghost{V_{BC}}&  \multigate{1}{V_{AC}} & \ghost{U_{BC}}& \multigate{1}{U_{AC}}&\qw\\
\lstick{A} & \qw & \ghost{V_{AC}} & \qw  &\ghost{U_{AC}}& \qw
}
\]
\end{lem}

\textit{Proof---}: Invoking Lemma 4.1, we only need to study the case that all the two-qubit gate are nonlocal.
The circuit is $$U_{AC}U_{BC}V_{AC}V_{BC}=V(\theta_1,\theta_2)_{ABC},$$ then $U_{AC}U_{BC}V_{AC}$ is a control unitary with control on $A$.
Moreover, for any input state $\ket{0}_A\ket{\psi}_{BC}$, the $A$'s part state of the following state is is $\ket{0}_A$.
$$U_{AC}U_{BC}V_{AC}~\ket{i}_A\ket{\psi}_{BC}.$$
After moving the local unitaries by invoking Proposition 1, we assume
$$V_{AC}\ket{0}_A\ket{0}_C=\ket{0}_A\ket{0}_C$$
Thus, the $A$'s part's state of the following state is $\ket{0}_A$,
\begin{eqnarray*}
U_{AC}U_{BC}V_{AC}\ket{0}_A\ket{y}_B\ket{0}_C=U_{AC}U_{BC}\ket{0}_A\ket{y}_B\ket{0}_C.
\end{eqnarray*}
There are three cases about the states $U_{BC}\ket{y}_B\ket{0}_C$:

Case 1:  There is some $\ket{y_0}_B$ such that $U_{BC}\ket{y}_B\ket{0}_C$ becomes entangled. Assume there is $0<\lambda<1$ such that
\[U_{BC}\ket{y_0}_B\ket{0}_C=\sqrt{\lambda}\ket{\alpha}_B\ket{0}_C+\sqrt{1-\lambda}\ket{\alpha^{\bot}}_B\ket{1}_C.\]
Define $\ket{\chi}_{ABC}=U_{AC}U_{BC}\ket{0}_A\ket{0}_B\ket{z_0}_C$, we know that
\begin{eqnarray*}
\ket{\chi}_{ABC}=\sqrt{\lambda}\ket{\Phi}_{AC}\ket{\alpha}_B+\sqrt{1-\lambda}\ket{\Psi}_{AC}\ket{\alpha^{\bot}}_B,\\
\Rightarrow\chi_A=\lambda\Phi_A+(1-\lambda)\Psi_A \Rightarrow\Phi_A=\Psi_A=\op{0}{0},
\end{eqnarray*}
where $\ket{\Phi}=U_{AC}\ket{00}_{AC}$ and $\ket{\Psi}=U_{AC}\ket{01}_{AC}$.
Therefore, $U_{AC}$ is a control unitary with control on system $A$, so is $V_{AC}$.
We can assume that
\begin{eqnarray*}
U_{AC}&=&\op{0}{0}\otimes I_C+\op{1}{1}\otimes w_{C_1},\\
V_{AC}&=&\op{0}{0}\otimes I_C+\op{1}{1}\otimes w_{C_2}.
\end{eqnarray*}
We know that $U_{BC}V_{BC}=I_{BC}$ and
\begin{eqnarray*}
w_{C_1}U_{BC}w_{C_2}V_{BC}=R_{BC} \Rightarrow U_{BC}w_{C_2}U_{BC}^{\dag}=w_{C_1}^{\dag}R_{BC}=w_{C_1}^{\dag}\oplus w_{C_1}^{\dag}D,
\end{eqnarray*}
where $D=\diag\{e^{i\theta_1},e^{i\theta_2}\}$ and
$$R=\op{00}{00}+\op{01}{01}+e^{i\theta_1}\op{10}{10}+e^{i\theta_2}\op{11}{11}.$$

$\{e^{i\varphi_1},e^{i\varphi_2},e^{i\varphi_1},e^{i\varphi_2}\}$ are the eigenvalues of $U_{BC}w_{C_2}U_{BC}^{\dag}$, where $e^{i\varphi_1},e^{i\varphi_2}$ are the eigenvalues of $w_{C_2}$. It is direct to see that $w_{C_1}^{\dag}$ can not have two identical eigenvalues, then $w_{C_1}^{\dag}$ and $w_{C_1}^{\dag}D$ enjoys the same eigenvalues, that leas us to that they have the same determinant.
$$\det(w_{C_1}^{\dag})=\det(w_{C_1}^{\dag}D)=\det(w_{C_1}^{\dag})\det(D)\Rightarrow \det(D)=1\Rightarrow e^{i(\theta_1+\theta_2)}= 1$$
Contradiction!

Otherwise, for any $\ket{y}_B$, $U_{BC}\ket{y}_B\ket{0}_C$ is product.

Case 2: There is a $\ket{\beta}_B$ and a local unitary $w_C$ on system $C$
such that $U_{BC}\ket{y}_B\ket{0}_C=\ket{\beta}_B w_C\ket{y}_C$,
thus, $U_{AC}$ maps $\{\ket{0}_A\}\otimes \mathcal{H}_C$ to itself.
Therefore, $U_{AC}$ is a control unitary with control on $A$, so is $V_{AC}$. The rest argument of this case is the same as case 1.

Case 3: There is a state on system $C$, wlog, says $\ket{0}_C$, and a local unitary $v_B$ on system $B$
such that $U_{BC}\ket{y}_B\ket{0}_C=v_B\ket{y}_B\ket{0}_C $.
Therefore, $U_{BC}$ is a control unitary with control on $C$. By moving this $v_B$ to $V_{BC}$, we make the assumption that
$$U_{BC}=\op{0}{0}\otimes I_B+\op{1}{1}\otimes u_B.$$
Note that for any $\ket{y}_B$, we have
\begin{eqnarray*}
V(\theta_1,\theta_2)_{ABC}\ket{0}_A (V_{BC}^{\dag}\ket{y}_B\ket{0}_C)&=&U_{AC}U_{BC}V_{AC}\ket{0}_A\ket{y}_B\ket{0}_C,\\
\Longrightarrow \ket{0}_A (V_{BC}^{\dag}\ket{y}_B\ket{0}_C)&=&U_{AC}\ket{0}_A\ket{y}_B\ket{0}_C.
\end{eqnarray*}
Thus part $B$'s state of $V_{BC}^{\dag}\ket{y}_B\ket{0}_C$ is $\ket{y}_B$ for all $\ket{y}_B\in\mathcal{H}_B$, which means that there is $\ket{\gamma}_C$ such that $V_{BC}\ket{y}_B\ket{\gamma}_C=\ket{y}_B\ket{0}_C.$
Therefore, one can find a unitary $w_C$ such that \[V_{BC}=\op{0}{\gamma}\otimes I_B+\op{1}{\gamma^{\bot}}\otimes v_B.\]
In order to simplify the structure of the two-qubit gates, we observe that
$$V_{BC}^{T}V_{AC}^{T}U_{BC}^{T}U_{AC}^{T}=V(\theta_1,\theta_2)_{ABC}^{T}=V(\theta_1,\theta_2)_{ABC},$$  hence also provides a simulation of $V(\theta_1,\theta_2)_{ABC}$.
Now we consider the state
$$V_{BC}^{T}V_{AC}^{T}U_{BC}^{T}\ket{x}_A\ket{0}_B\ket{0}_C=V_{BC}^{T}V_{AC}^{T}\ket{x}_A\ket{0}_B\ket{0}_C$$ for any $\ket{x}_A$.
The argument of cases 1 and 2 excludes the following possibilities:
(i) there is some $\ket{x}_A$ such that $V_{AC}^{T}\ket{x}_A\ket{0}_C$ is entangled, or (ii)
there is a $\ket{\delta}_A$ and a local unitary $w_C$ on system $C$ such that $V_{AC}^{T}\ket{x}_A\ket{0}_C=\ket{\delta}_Aw_C\ket{x}_C$.

So the only possibility is that there is a state $\ket{\phi}_C$ on system $C$, and a local unitary $w_A$
on system $A$ such that $V_{AC}^{T}\ket{x}_A\ket{0}_C=w_A\ket{x}_A\ket{\phi}_C $.

According to $V_{AC}\ket{0}_A\ket{0}_C=\ket{0}_A\ket{0}_C$, we can choose $\ket{\phi}=\ket{0}$.
Thus $V_{AC}$ is a controlled gate with control system $C$, $i.e.$,
 $$V_{AC}=\op{0}{0}\otimes w_A+\op{1}{1}\otimes v_A.$$

By studying part $B$'s state of $$U_{AC}U_{BC}V_{AC}V_{BC}\ket{0}_A\ket{y}_{B}\ket{0}_C=\ket{0}_A\ket{y}_{B}\ket{0}_C,$$
we see that $\ket{\gamma}_C$ defined in $V_{BC}$ equals to $\ket{0}_B$ or $\ket{1}_B$,
up to some global phase. Otherwise, assume that $\ket{0}_C=a\ket{\gamma}_C+b\ket{\gamma^{\perp}}_C$ for $ab\neq 0$.
Then the state of part $B$ becomes a mixed state for general input $\ket{0}_A\ket{y}_{B}\ket{0}_C$ since $u_B$ is not identity up to some global phase and $U_{BC}$ is nonlocal.
For the case $\ket{\gamma}_C=\ket{0}_C$,
we know that all the four two-qubit gates are controlled gate with control system $C$, which implies that
$R_{BC}$ defined in case 1 is a local unitary, a contradiction.
For the case $\ket{\gamma}_C=\ket{1}_C$, let $X_C$ be the NOT (flip) gate such that $X\ket{0}=\ket{1}$ and $X\ket{1}=\ket{0}$, then one can verify that
\[(U_{AC}X_C)(X_CU_{BC}X_C)(X_CV_{AC}X_C)(X_CV_{BC})=V(\theta_1,\theta_2)_{ABC}.\]
Then $U_{AC}X_C,X_CU_{BC}X_C,X_CV_{AC}X_C$ and $X_CV_{BC}$
are all controlled gate with control system $C$.
This also leads us to the impossible conclusion that $R_{BC}$ is local. Impossible!

Now we are able to show that
\begin{thm}
$V(\theta_1,\theta_2)$ requires five two-qubit gates to simulate, provide that $e^{i(\theta_1+\theta_2)}\neq 1$ and $e^{i\theta_1}\neq e^{i\theta_2}$.
\end{thm}

\textit{Proof---}: If the four gates belong to two of the classes $\mathcal{K}_{AB}, \mathcal{K}_{AC}, \mathcal{K}_{BC}$, the circuits that need to be considered are just $U_{AC}U_{BC}V_{AC}V_{BC}=V(\theta_1,\theta_2)_{ABC}$ and $U_{AB}U_{BC}V_{AB}V_{BC}=V(\theta_1,\theta_2)_{ABC}$. The previous one is studied in Lemma 4.2. The latter one is impossible by using $V(0,\theta_2-\theta_1)_{ABC}=W(-\theta_1)_{AB}V(\theta_1,\theta_2)_{ABC}$ and Lemma 4.1 directly.

Otherwise, the four gates belong to three of the classes $\mathcal{K}_{AB}, \mathcal{K}_{AC}, \mathcal{K}_{BC}$, then there exist two gates belongs to the same class. Due to the symmetric property of the Fredkin gate, we need to consider the following two cases:

Case 1: Two gates belong to $\mathcal{K}_{AB}$, then at least one of them lies in the front or the end of the circuit.
then we conclude that is impossible by apply Lemma 4.1 and $$V(0,\theta_2-\theta_1)_{ABC}=W(-\theta_1)_{AB}V(\theta_1,\theta_2)_{ABC}=V(\theta_1,\theta_2)_{ABC}W(-\theta_1)_{AB}.$$

Case 2: Two gates belong to $\mathcal{K}_{BC}$. The four gates are $U_{AB}\in\mathcal{K}_{AB}$, $U_{AC}\in\mathcal{K}_{AC}$ and $U_{BC},V_{BC}\in\mathcal{K}_{BC}$. According to symmetric properties of the Fredkin gate, the three possible circuits are

Subcase 1: $U_{BC}U_{AC}V_{BC}U_{AB}=V(\theta_1,\theta_2)_{ABC}$, this is impossible since $U_{AB}$ lies in the end.

Subcase 2: $U_{BC}U_{AB}V_{BC}U_{AC}=F_{ABC}$.
this circuit can be reduced to Lemma 4.2 by noticing that $S_{BC}U_{BC},S_{BC}V_{BC}\in \mathcal{K}_{BC}$ and $S_{BC}U_{AC}S_{BC}\in \mathcal{K}_{AB}$ and $S_{BC}F_{ABC}S_{BC}$ is the controlled controlled gate with control on systems $A$ and $C$,
\[(S_{BC}U_{BC})U_{AB}(S_{BC}V_{BC})(S_{BC}U_{AC}S_{BC})=S_{BC}F_{ABC}S_{BC}.\]
Subcase 3: $U_{BC}U_{AB}U_{AC}V_{BC}=V(\theta_1,\theta_2)_{ABC}$. Observe that $U_{AB}U_{AC}$ is a control unitary with control on $A$, we can conclude that $R_{BC}$ share eigenvalues with a local unitary by invoking Proposition 2. That is impossible.

\section{Optimal simulation of Fredkin gate }

The Fredkin gate is the three-qubit gate that swaps the last two-qubits if the first-qubit is $\ket{1}$. The matrix form of Fredkin gate is given as
\begin{eqnarray*}
F=\left(
\begin{array}{cccccccc}
 1 & 0 & 0 & 0 & 0 & 0 & 0& 0\\
 0 & 1 & 0 & 0 & 0 & 0 & 0& 0\\
 0 & 0 & 1 & 0 & 0 & 0 & 0& 0\\
 0 & 0 & 0 & 1 & 0 & 0 & 0& 0\\
 0 & 0 & 0 & 0 & 1 & 0 & 0& 0\\
 0 & 0 & 0 & 0 & 0 & 0 & 1& 0\\
 0 & 0 & 0 & 0 & 0 & 1 & 0& 0\\
 0 & 0 & 0 & 0 & 0 & 0 & 0& 1\\
\end{array}
\right).
\end{eqnarray*}
It is showed that the Fredkin gate can be simulated by employing five two-qubit gate in the following circuit \cite{SD96} by letting $V^2=X$ with $X$ being the pauli flip matrix,
 
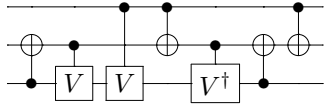
\[
\Qcircuit @C=.5em @R=0em @!R {
 & \qw & \qw & \ctrl{2} & \ctrl{1} & \qw & \qw &\ctrl{1} & \qw\\
 &\targ & \ctrl{1} & \qw & \targ & \ctrl{1} & \targ &\targ & \qw\\
 & \ctrl{-1}& \gate{V} & \gate{V} & \qw & \gate{V^\dag} &\ctrl{-1} &\qw & \qw
}
\]
In this section, we will show that
\begin{thm}
Four two-qubit gates are not sufficient for implementing $F_{ABC}$.
\end{thm}
The proof of Theorem 1 heavily depends on the discussion of the possible circuit structures. The following symmetric properties of the Fredkin gate are quite helpful to decrease the number of cases: the Fredkin gate is invariant under the permutation of $B$ and $C$ and it is symmetric, $i.e.$, $F_{ABC}=F_{ACB}$ and $F_{ABC}=F_{ABC}^{T}$ with $F_{ACB}=S_{BC}F_{ABC}S_{BC}$.

$F_{ABC}$ can be regarded as a tripartite unitary of Hilbert space $\mathcal{H}_A\otimes \mathcal{H}_B\otimes \mathcal{H}_C$. We can easily verify $F_{ABC}$ is a control unitary with control on $A$ by noticing
\[F_{ABC}=\op{0}{0}\otimes I_{BC}+\op{1}{1}\otimes S_{BC},\]
where $I$ and $S$ stand for an identity operator and the swap gate, respectively, the index $BC$ means the system it applies on is $\mathcal{H}_B\otimes \mathcal{H}_C$, a qubit-qudit unitary $U$ is called a control unitary if there exist unitaries $U_0$ and $U_1$ such that $U=\op{0}{0}\otimes U_0+\op{1}{1}\otimes U_1$.

In order to explain our idea and key technique of showing four gates are not enough, we first demonstrate that a Fredkin gate can not be decomposed into two two-qubit gates by dividing the problem into two cases: Case 1: The two gates belong to classes $\mathcal{K}_{AB}, \mathcal{K}_{BC}$, then $U_{AB}$ must be a control unitary with control on $A$, A direct calculation leads to the confliction; Case 2: The two gates belong to two of the classes $\mathcal{K}_{AB}, \mathcal{K}_{AC}$, we assume the circuit is
$U_{AB}U_{AC}=F_{ABC}$, invoking Proposition 2, we can assert that $S_{BC}$ is a local unitary by figuring directly out the form of control unitary. That is again impossible. Therefore,
\begin{lem}
Two two-qubit gates are not sufficient for implementing $F_{ABC}$.
\end{lem}

In the rest, we show that four nonlocal two-qubit gates are not sufficient for implementing $F_{ABC}$.
We first study two special kinds of circuits.
\begin{lem}
There is no $U_{AB},V_{AB}\in \mathcal{K}_{AB}$ and $U_{BC},V_{BC}\in \mathcal{K}_{BC}$ such that the Fredkin gate can be implemented in the following circuit,
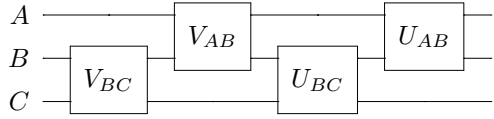
\[
\Qcircuit @C=1em @R=.7em {
\lstick{A} & \qw                  &  \multigate{1}{V_{AB}} & \qw                    & \multigate{1}{U_{AB}} & \qw \\
\lstick{B} & \multigate{1}{V_{BC}}&  \ghost{V_{AB}}        &  \multigate{1}{U_{BC}} & \ghost{U_{AB}}&\qw\\
\lstick{C} & \ghost{V_{BC}}       & \qw                    &  \ghost{U_{BC}}        & \qw &\qw
}
\].
\end{lem}
\textit{Proof---}: The circuit is $U_{AB}U_{BC}V_{AB}V_{BC}=F_{ABC}$, then $U_{AB}U_{BC}V_{AB}$ is a control unitary with control on $A$.
Moreover, for any input state $\ket{0}_A\ket{\psi}_{BC}$, the $A$'s part state of
$U_{AB}U_{BC}V_{AB}~\ket{i}_A\ket{\psi}_{BC}$
is $\ket{0}_A$.
We assume
$V_{AB}\ket{0}_A\ket{0}_B=\ket{0}_A\ket{0}_B$ after moving the local unitaries by invoking Proposition 1.
Thus, the $A$'s part's state of the following state is $\ket{0}_A$,
\begin{eqnarray*}
U_{AB}U_{BC}V_{AB}\ket{0}_A\ket{0}_B\ket{z}_C=U_{AB}U_{BC}\ket{0}_A\ket{0}_B\ket{z}_C.
\end{eqnarray*}
There are three cases about the states $U_{BC}\ket{0}_B\ket{y}_C$:

Case 1:  There is some $\ket{z_0}_C$ such that $U_{BC}\ket{0}_B\ket{z_0}_C$ becomes entangled. Assume there is $0<\lambda<1$ such that
\[U_{BC}\ket{0}_B\ket{z_0}_C=\sqrt{\lambda}\ket{0}_B\ket{\alpha}_C+\sqrt{1-\lambda}\ket{1}_B\ket{\alpha^{\bot}}_C.\]
Define $\ket{\chi}_{ABC}=U_{AB}U_{BC}\ket{0}_A\ket{0}_B\ket{z_0}_C$, we know that
\begin{eqnarray*}
\ket{\chi}_{ABC}=\sqrt{\lambda}\ket{\Phi}_{AB}\ket{\alpha}_C+\sqrt{1-\lambda}\ket{\Psi}_{AB}\ket{\alpha^{\bot}}_C,\\
\Rightarrow\chi_A=\lambda\Phi_A+(1-\lambda)\Psi_A \Rightarrow\Phi_A=\Psi_A=\op{0}{0},
\end{eqnarray*}
where $\ket{\Phi}=U_{AB}\ket{00}$ and $\ket{\Psi}=U_{AB}\ket{01}$.
Therefore, $U_{AB}$ is a control unitary with control on system $A$, so is $V_{AB}$.
We can assume that
\begin{eqnarray*}
U_{AB}&=&\op{0}{0}\otimes I_B+\op{1}{1}\otimes v_{B_1},\\
V_{AB}&=&\op{0}{0}\otimes I_B+\op{1}{1}\otimes v_{B_2}.
\end{eqnarray*}
We know that $U_{BC}V_{BC}=I_{BC}$ and
\begin{eqnarray*}
v_{B_1}U_{BC}v_{B_2}V_{BC}=S_{BC} \Rightarrow U_{BC}v_{B_2}U_{BC}^{\dag}=v_{B_1}^{\dag}S_{BC}.
\end{eqnarray*}
$\{e^{i\theta_1},e^{i\theta_2},e^{i(\theta_1+\theta_2)/2},-e^{i(\theta_1+\theta_2)/2}\}$ are the eigenvalues of $v_{B_1}^{\dag}S_{BC}$, where $e^{i\theta_1},e^{i\theta_2}$ are the eigenvalues of $v_{B_1}^{\dag}$. One can verify that these can not be the eigenvalues of $v_{B_2}\otimes I_C$, contradict to the above equation.

Otherwise, for any $\ket{z}_C$, $U_{BC}\ket{0}_B\ket{z}_C$ is product.

Case 2: There is a $\ket{\gamma}_C$ and a local unitary $w_B$ on system $B$
such that $U_{BC}\ket{0}_B\ket{z}_C=w_B\ket{z}_B\ket{\gamma}_C$,
thus, $U_{AB}$ maps $\{\ket{0}_A\}\otimes \mathcal{H}_B$ to itself.
Therefore, $U_{AB}$ is a control unitary with control on $A$, so is $V_{AB}$. The rest argument of this case is the same as case 1.

Case 3: There is a state on system $B$, wlog, says $\ket{0}_B$, and a local unitary $w_C$ on system $C$
such that $U_{BC}\ket{0}_B\ket{z}_C=\ket{0}_B w_C\ket{z}_C$.
Therefore, $U_{BC}$ is a control unitary with control on $B$. By moving this $w_C$ to $V_{BC}$, we make the assumption that
$U_{BC}=\op{0}{0}\otimes I_C+\op{1}{1}\otimes u_C.$
Note that for any $\ket{z}_C$, we have
\begin{eqnarray*}
F_{ABC}\ket{0}_A (V_{BC}^{\dag}\ket{0}_{B}\ket{z}_C)&=&U_{AB}U_{BC}V_{AB}\ket{0}_A\ket{0}_{B}\ket{z}_C\\
\Longrightarrow \ket{0}_A (V_{BC}^{\dag}\ket{0}_{B}\ket{z}_C)&=&U_{AB}\ket{0}_A\ket{0}_{B}\ket{z}_C
\end{eqnarray*}
Thus part $C$'s state of $V_{BC}^{\dag}\ket{0}_{B}\ket{z}_C$ is $\ket{z}_C$ for all $\ket{z}_C\in\mathcal{H}_C$, which means that there is $\ket{\beta}_B$ such that $V_{BC}\ket{\beta}_B\ket{z}_C=\ket{0}_B\ket{z}_C.$
Therefore, one can find a unitary $w_C$ such that \[V_{BC}=\op{0}{\beta}\otimes I_C+\op{1}{\beta^{\bot}}\otimes w_C.\]
The state of $C'$s part of the following state is always $\ket{\beta}_B$,
\[U_{BC}V_{AB}V_{BC}\ket{1}_A\ket{\beta}_B\ket{z}_C=U_{BC}V_{AB}\ket{1}_A\ket{0}_B\ket{z}_C.\]
On the other hand, direct calculation shows that the state of $C'$s part of the above state is the mixture of $\ket{z}_C$ and $w_C\ket{z}_C$, which is not constant state. Conflict!

\begin{lem}
There is no $U_{AB},V_{AB}\in \mathcal{K}_{AB}$ and $U_{AC},V_{AC}\in \mathcal{K}_{AC}$ such that $U_{AC}U_{AB}V_{AC}V_{AB}=F_{ABC}$, $i.e.$, the Fredkin gate can be implemented in the following circuit,
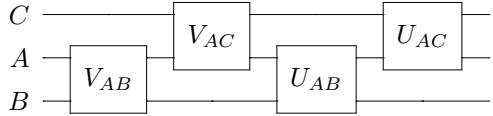
\[
\Qcircuit @C=1em @R=.7em {
\lstick{C} & \qw                  &  \multigate{1}{V_{AC}} & \qw                    & \multigate{1}{U_{AC}} & \qw \\
\lstick{A} & \multigate{1}{V_{AB}}&  \ghost{V_{AC}}        &  \multigate{1}{U_{AB}} & \ghost{U_{AC}}&\qw\\
\lstick{B} & \ghost{V_{AB}}       & \qw                    &  \ghost{U_{AB}}        & \qw &\qw
}
\].
\end{lem}
\textit{Proof---}: Wolg, we assume that
$V_{AB}\ket{0}_A\ket{\beta}_B=\ket{0}_A\ket{0}_B$ by invoking Proposition 1.
Then we have the following equation
\begin{eqnarray*}
U_{AC}U_{AB}V_{AC}V_{AB}\ket{0}_A\ket{\beta}_B\ket{z}_C=F_{ABC}\ket{0}_A\ket{\beta}_B\ket{z}_C.
\end{eqnarray*}
Considering the states with form $V_{AC}\ket{0}_A\ket{z}_C$:\\
Case 1:  There is some $\ket{z_0}_C$ such that $V_{AC}\ket{0}_A\ket{z_0}_C$ becomes entangled. Define $\ket{\chi}_{ABC}$ as follows,
\[\ket{\chi}_{ABC}:=U_{AC}U_{AB}V_{AC}\ket{0}_A\ket{0}_B\ket{z_0}_C=\ket{0}_A\ket{\beta}_B\ket{z_0}_C.\]
We can assume there is $0<\lambda<1$ such that
\[V_{AC}\ket{0}_A\ket{z_0}_C=\sqrt{\lambda}\ket{0}_A\ket{0}_C+\sqrt{1-\lambda}\ket{1}_A\ket{1}_C.\]
According to the fact that $\chi_B=\beta_B$, we know that
\[U_{AB}=I_A\otimes\op{\beta}{0}+u_A\otimes\op{\beta^{\bot}}{1}.\]
One can verify that $\omega_B=\beta^{\bot}$ by noticing that $U_{AC}\ket{\omega}_{ABC}=F_{ABC}\ket{0}_A\ket{\beta^{\bot}}_B\ket{z}_C=\ket{0}_A\ket{\beta^{\bot}}_B\ket{z}_C$, where
$\ket{\omega}_{ABC}=U_{AB}V_{AC}V_{AB}\ket{0}_A\ket{\beta^{\bot}}_B\ket{z}_C$.

Employe the form of $U_{AB}$ and $\omega_B=\beta^{\bot}$, we know that $V_{AB}\ket{0}_A\ket{\beta^{\bot}}_B=\ket{\phi}_A\ket{1}_B$ for some $\ket{\phi}_A\in \mathcal{H}_A$.

Notice that $\ket{1}_A\ket{0}_B$ is orthogonal to $V_{AB}\ket{0}_A\ket{\beta^{\bot}}_B=\ket{\phi}_A\ket{1}_B$ and $V_{AB}\ket{0}_A\ket{\beta}_B=\ket{0}_A\ket{0}_B$, then there is a $\ket{\xi}_B\in\mathcal{H}_B$ such that $V_{AB}\ket{1}_A\ket{\xi}_B=\ket{1}_A\ket{0}_B$ since $V_{AB}$ is a unitary. Now we consider the states with form
\begin{eqnarray*}
\ket{\psi}_{ABC}=U_{AB}V_{AC}\ket{1}_A\ket{0}_B\ket{z}_C.
\end{eqnarray*}
Then we know that $\psi_B=\op{z}{z}$ by noticing that \[\ket{\psi}_{ABC}=U_{AC}^{\dag}F_{ABC}\ket{1}_A\ket{\xi}_B\ket{z}_C=U_{AC}^{\dag}\ket{1}_A\ket{z}_B\ket{\xi}_C.\]
We can observe that $\psi_B=\op{\beta}{\beta}$ by employing the form of $U_{AB}$. Conflict!

Otherwise, for any $\ket{z}_C$, $V_{AC}\ket{0}_A\ket{z}_C$ is product.\\
Case 2: There is a $\ket{\gamma}_C$ and a local unitary $u_A$ on system $A$
such that $V_{AC}\ket{0}_A\ket{z}_C=u_A\ket{z}_A\ket{\gamma}_C$, thus, $U_{AB}$ maps $\mathcal{H}_A\otimes \{\ket{0}_B\}$ to $\mathcal{H}_A\otimes \{\ket{\beta}_B\}$. We can also obtain the form of $U_{AB}$ as case 1. Repeating the argument of case 1, we are able to show the impossibility of this case.\\
Case 3: There is a state $\ket{\alpha}_A$ on system $A$ such that $V_{AC}$ maps $\{\ket{0}_A\otimes \mathcal{H}_C\}$ to $\{\ket{\alpha}_A\otimes \mathcal{H}_C\}$. Moving the local unitary, we can make $\ket{\alpha}_A=\ket{0}_A$, that means $V_{AC}$ is a control unitary with control on $A$. We can assume that $V_{AC}=\op{0}{0}\otimes I_C+\op{1}{1}\otimes w_C$ by moving the local unitary, then $\ket{0}_A\ket{\beta}_B\ket{z}_C$ equals to
\begin{eqnarray*}
U_{AC}U_{AB}V_{AC}V_{AB}\ket{0}_A\ket{\beta}_B\ket{z}_C=U_{AC}U_{AB}\ket{0}_A\ket{0}_B\ket{z}_C.
\end{eqnarray*}
By moving the local unitary to the left of $U_{AC}$, we can make the assumption $U_{AB}\ket{0}_A\ket{0}_B=\ket{0}_A\ket{\beta}_B$, then
\begin{eqnarray*}
\ket{0}_A\ket{z}_C=U_{AC}\ket{0}_A\ket{z}_C,
\end{eqnarray*}
that is $U_{AC}$ is a control unitary with control on $A$.

Since $F_{ABC}$ is symmetric, we observe that
\[V_{AB}^{T}V_{AC}^{T}U_{AB}^{T}U_{AC}^{T}=F_{ABC}^{T}=F_{ABC}.\]
Repeating the argument above, we can derive the confliction if there is some $\ket{y}_B$ such that $U_{AB}^{T}\ket{0}_A\ket{y}_B$ becomes entangled or some $\ket{\varphi}_C$ and some local unitary $u_{A1}$ on system $A$
such that $U_{AB}^{T}\ket{0}_A\ket{y}_B=u_{A1}\ket{y}_A\ket{\varphi}_C$ is valid for any $\ket{y}_B$. Otherwise, we can reach the conclusion that $V_{AB}^{T}$ and $U_{AB}^{T}$ are both control unitary with control on $A$, so are $V_{AB}$ and $U_{AB}$, we can assert that $S_{BC}$ is a local unitary by figuring directly out the form of control unitary. That is again impossible.

Now we are ready to show that
\begin{thm}
Four nonlocal two-qubit gates are not sufficient for implementing a Fredkin gate.
\end{thm}
\textit{Proof---}: If the four gates belong to two of the classes $\mathcal{K}_{AB}, \mathcal{K}_{AC}, \mathcal{K}_{BC}$, the circuits that need to be considered is just $U_{AB}U_{BC}V_{AB}V_{BC}=F_{ABC}$ or $U_{AB}U_{AC}V_{AB}V_{AC}=F_{ABC}$, which just have been studied in Lemma 5.2 and 5.3.

Otherwise, the four gates belong to three of the classes $\mathcal{K}_{AB}, \mathcal{K}_{AC}, \mathcal{K}_{BC}$, then there exist two gates belongs to the same class. Due to the symmetric property of the Fredkin gate, we need to consider the following two cases:\\
Case 1: Two gates belong to $\mathcal{K}_{AB}$. The four gates are $U_{AB},V_{AB}\in\mathcal{K}_{AB}$, $U_{AC}\in\mathcal{K}_{AC}$ and $U_{BC}\in\mathcal{K}_{BC}$. According to symmetric properties of the Fredkin gate, the three possible circuits are\\
Subcase 1: $U_{AB}U_{BC}V_{AB}U_{AC}=F_{ABC}$. This circuit can be reduced to Lemma 3 by noticing that
\[(S_{AB}U_{AB})U_{BC}(V_{AB}S_{AB})(S_{AB}U_{AC}S_{AB})=F_{BAC},\]
$S_{AB}U_{AB}, V_{AB}S_{AB}\in \mathcal{K}_{AB}$ and $S_{AB}U_{AC}S_{AB}\in \mathcal{K}_{BC}$, where $F_{BAC}=S_{AB}F_{ABC}S_{AB}$ is the Fredkin gate with control on $B$. \\
Subcase 2: $U_{AB}U_{AC}V_{AB}U_{BC}=F_{ABC}$. This circuit can be reduced to Lemma 2 by noticing that
\[(S_{AB}U_{AB})U_{AC}(V_{AB}S_{AB})(S_{AB}U_{BC}S_{AB})=F_{BAC},\]
and $S_{AB}U_{BC}S_{AB}\in \mathcal{K}_{AC}$.\\
Subcase 3: $U_{AB}U_{BC}U_{AC}V_{AB}=F_{ABC}$. This circuit can be reduced to subcase 1 by noticing that
\[U_{AB}(U_{BC}S_{BC})(S_{BC}U_{AC}S_{BC})(S_{BC}V_{AB}S_{BC})=F_{ABC}S_{BC}.\]
$S_{BC}U_{AC}S_{BC}\in\mathcal{K}_{AB}$ and $S_{BC}V_{AB}S_{BC}\in\mathcal{K}_{AC}$.
Observe that $X_AF_{ABC}S_{BC}X_A=F_{ABC}$ where $X_A$ denotes the Pauli flip unitary on system $\mathcal{H}_A$. This reduction is done.\\
Case 2: Two gates belong to $\mathcal{K}_{BC}$. The four gates are $U_{AB}\in\mathcal{K}_{AB}$, $U_{AC}\in\mathcal{K}_{AC}$ and $U_{BC},V_{BC}\in\mathcal{K}_{BC}$. According to symmetric properties of the Fredkin gate, the three possible circuits are\\
Subcase 1: $U_{BC}U_{AC}V_{BC}U_{AB}=F_{ABC}$.  This circuit can be reduced to Lemma 2 by noticing that
\[(S_{BC}U_{BC})U_{AC}(V_{BC}S_{BC})(S_{BC}U_{AB}S_{BC})=F_{ABC}.\]
Subcase 2: $U_{BC}U_{AB}V_{BC}U_{AC}=F_{ABC}$.
his circuit can be reduced to Lemma 3 by noticing that
\[(S_{BC}U_{BC})U_{AB}(S_{BC}V_{BC})(S_{BC}U_{AC}S_{BC})=F_{ABC}.\]
Subcase 3: $U_{BC}U_{AB}U_{AC}V_{BC}=F_{ABC}$. Observe that $U_{AB}U_{AC}$ is a control unitary with control on $A$, we can conclude that $S_{BC}$ share eigenvalues with a local unitary by invoking Proposition 2. That is impossible.

This completes the proof of this theorem.

Now we are safe to say that the Fredkin gate can not be simulated by using four two-qubit gates. Together with the previous known result, we can conclude that five two-qubit gates is optimal to implement a Fredkin gate.

By observing that the set of circuits consisting of four two-qubit gates forms a closed set, compact set indeed, a direct corollary of Theorem 1 is that
\begin{cor}
There is $\epsilon>0$ such that for any $U_{ABC}$ which could be implemented by four two-qubit gates, the distance between $U_{ABC}$ and $F_{ABC}$ is greater than $\epsilon$. In other words, Fredkin gate can not be well approximated by any circuit consisting of four two-qubit gates.
\end{cor}

\section{Conclusion.}

In this paper, we study the problem of implementing a three-qubit controlled controlled gate and Fredkin gate using two-qubit unitaries. We first showed that any controlled controlled gate requires at least four two-qubit gates to implement, and if the determinant is one then it can be simulated using four two-qubit gates, otherwise, five is optimal. We can construct a set of universal controlled controlled gate which can be simulated by a circuit consisting of four two-qubit gates.

Secondly, we proved that five two-qubit gates is optimal for constructing a three-qubit Fredkin gate. We hope this work will be helpful to study the problem that whether six two-bit quantum gates
are sufficient to generate any three-bit quantum gate.

This work was partly supported by the National Natural Science Foundation
of China(Grant Nos. 61179030 and 60621062), the Australian Research
Council (Grant Nos. DP110103473 and DP120103776) and the Overseas Team Program of Academy of Mathematics and
Systems Science, Chinese Academy of Sciences.

\end{document}